\documentclass[fleqn,usenatbib]{mnras}

\pdfoutput=1
\usepackage{graphicx,subfigure,bm,color,psfrag,hyperref}
\usepackage{amsfonts}
\usepackage{lipsum}
\usepackage{mathtools}
\usepackage{verbatim}
\usepackage{color}
\usepackage{newtxtext,newtxmath}
\usepackage[T1]{fontenc}
\usepackage{amsmath} 

\DeclareRobustCommand{\VAN}[3]{#2}
\let\VANthebibliography\thebibliography
\def\thebibliography{\DeclareRobustCommand{\VAN}[3]{##3}\VANthebibliography}

\begin{document}

\title[A fake Interacting Dark Energy detection?]{A fake Interacting Dark Energy detection?}

\author[Eleonora Di Valentino and Olga Mena]{
Eleonora Di Valentino $^{1}$\thanks{E-mail: eleonora.divalentino@manchester.ac.uk}
and Olga Mena $^{2}$ \\
$^{1}$Jodrell Bank Center for Astrophysics, School of Physics and Astronomy, 
University of Manchester, Oxford Road, Manchester, M13 9PL, UK\\
$^{2}$IFIC, Universidad de Valencia-CSIC, 46071, Valencia, Spain
}

\date{Accepted XXX. Received YYY; in original form ZZZ}

\pubyear{2020}

\label{firstpage}
\pagerange{\pageref{firstpage}--\pageref{lastpage}}
\maketitle
\begin{abstract}
Models involving an interaction between the Dark Matter and the Dark Energy sectors have been proposed to alleviate the long standing Hubble constant tension. In this paper we analyze whether the constraints and potential hints obtained for these interacting models remain unchanged when using simulated Planck data. Interestingly, our simulations indicate that a dangerous fake detection for a non-zero interaction among the Dark Matter and the Dark Energy fluids could arise when dealing with current CMB Planck measurements alone. The very same hypothesis is tested against future CMB observations, finding that only cosmic variance limited polarization experiments, such as PICO or PRISM, could be able to break the existing parameter degeneracies and provide reliable cosmological constraints. This paper underlines the extreme importance of confronting the results arising from data analyses with those obtained with simulations when extracting cosmological limits within exotic cosmological scenarios.

\end{abstract}

\section{Introduction}

Despite the wonderful agreement of present cosmological measurements with the canonical $\Lambda$CDM model, some tensions between different observations have started to question the validity of the standard cosmological model. In particular, a significant role is the one played by the long standing Hubble constant tension at $4.4\sigma$ between the estimate from 2018 Planck released data~\cite{Aghanim:2018eyx} and the value measured by R19, i.e. the SH0ES collaboration~\cite{Riess:2019cxk} (see~\cite{DiValentino:2020zio} for a recent overview). These two recent  measurements are supported by other early and late time cosmological probes, respectively, making more difficult the possibility of isolating systematic effects in the experiments that could bias the data in the same direction. For this reason, a gigantic effort has been made by the community to build models beyond the standard $\Lambda$CDM that could explain and alleviate the $H_0$ disagreement with a modification of the cosmological scenario.

Plenty of work has been devoted to early modifications of the expansion history, such adding either an Early Dark Energy component~\cite{Pettorino:2013ia,Poulin:2018cxd,Karwal:2016vyq,Sakstein:2019fmf,Niedermann:2019olb,Akarsu:2019hmw,Ye:2020btb,Agrawal:2019lmo,Lin:2019qug,Berghaus:2019cls,Smith:2019ihp,Lucca:2020fgp} or extra relativistic species at recombination~\cite{Anchordoqui:2011nh,Jacques:2013xr,Weinberg:2013kea,Anchordoqui:2012qu,Carneiro:2018xwq,Paul:2018njm,DiValentino:2015sam,Green:2019glg,Ferreira:2018vjj,Gelmini:2019deq,DiValentino:2015wba,Poulin:2018dzj,Baumann:2016wac,Barenboim:2016lxv,Zeng:2018pcv,Allahverdi:2014ppa}. Interestingly, these solutions are promising for both solving the $H_0$ tension and lowering the sound horizon at the drag epoch~\cite{Knox:2019rjx,Evslin:2017qdn}, despite the fact that they do not provide a value for the Hubble constant large enough to be in agreement with R19~\cite{Arendse:2019hev}. On the other hand, late time modifications of the expansion history, such as the phantom Dark Energy~\cite{Aghanim:2018eyx,Yang:2018qmz,Yang:2018prh,DiValentino:2019dzu,Vagnozzi:2019ezj,DiValentino:2020naf,Keeley:2019esp,Joudaki:2016kym} or the Phenomenologically Emergent Dark Energy~\cite{Li:2019yem,Pan:2019hac,Rezaei:2020mrj,Liu:2020vgn,Li:2020ybr,Yang:2020tax} scenarios, perform better in the $H_0$ resolution but leave the sound horizon unaltered.

Within this context, a large number of the models proposed involves the encouraging possibility of an interaction between Dark Matter and Dark Energy (IDE models~\cite{Pettorino:2013oxa,Salvatelli:2014zta,Kumar:2016zpg,DiValentino:2017iww,Kumar:2017dnp,vandeBruck:2017idm,Sola:2017znb,Yang:2018euj,Yang:2018uae,Yang:2019uzo,Martinelli:2019dau, DiValentino:2019ffd,DiValentino:2019jae,Benevento:2020fev,Gomez-Valent:2020mqn,Lucca:2020zjb,Yang:2020uga,Yang:2019uog,Yang:2018ubt,Agrawal:2019dlm,Anchordoqui:2019amx,Johnson:2020gzn,Anchordoqui:2020sqo}). This solution naturally releases the Hubble constant tension, because it plays with the geometrical degeneracy existing between the parameter governing the interacting rate and the dark matter mass energy density, which is modified by the flux of energy exchanged between the dark matter and the dark energy fluids. 

In this paper we scrutinize if the current constraints obtained for these IDE models are reliable simulating the Planck data, and if future CMB experiments can improve the soundness of present IDE bounds. We present in Sec.~\ref{Method} the approach used in this paper to simulate and analyse the mock datasets. Section~\ref{results} contains our main results. We conclude in Sec.~\ref{conclu}.

\section{Methodology}
\label{Method}

\begin{table}
	\centering
	\caption{Experimental specifications.}
	\label{tab:specifications}
\begin{tabular}{|c|c|c|c|c|c|c|}
\hline
Configuration                    &  Channel& Beam &$\Delta P$& $\ell_{max}$&$\ell_{min}$& $f_{\rm sky}$\\
& GHz& arcmin &$\mu K$-arcmin& && \\
\hline
PICO  &$75$   & $10.7$ & $4.2$ & $4000$&$2$& $0.75$\\
   &$90$ & $9.5$ & $2.8$ & && \\
  &$108$& $7.9$ & $2.3$ & && \\
 &$129$ & $7.4$ & $2.1$ & && \\
  &$155$  & $6.2$ & $1.8$ & && \\
&$186$ & $4.3$ & $4.0$ & && \\
 &$223$ & $3.6$ & $4.5$ & && \\
\hline
PRISM    &$52$   & $7.35$ & $6.08$ & $6000$&$2$& $0.75$\\
&$62$  & $5.12$ & $5.86$ & && \\
&$75$  & $4.27$ & $4.56$ & && \\
&$90$  & $3.80$ & $3.04$ & && \\
&$108$  & $3.16$ & $2.39$ & && \\
&$129$  & $2.96$ & $2.39$ & && \\
&$155$  & $2.48$ & $1.95$ & && \\
&$186$  & $1.72$ & $4.34$ & && \\
&$223$  & $1.44$ & $4.99$ & && \\
&$268$  & $1.28$ & $3.26$ & && \\
&$321$  & $1.04$ & $4.56$ & && \\
&$385$  & $1.00$ & $4.99$ & && \\
\hline
\end{tabular}
\end{table}

For simulating current and future CMB measurements, we shall follow the approach used in several white papers and commonly exploited in the literature, see e.g. Refs.~\cite{DiValentino:2016foa,Hanany:2019lle,Delabrouille:2019thj,Capparelli:2017tyx,DiValentino:2018jbh,Renzi:2018dbq,Renzi:2017cbg,DiValentino:2019qzk}. The fiducial cosmology is a vanilla flat $\Lambda$CDM model compatible with Planck TT,TE,EE + lowE measurements, and therefore with a null coupling between Dark Matter and Dark Energy. The values of the parameters assumed for our simulated data are reported in the second column of the Tables~\ref{table_IDE} and~\ref{table_LCDM}.

We compute the theoretical CMB angular power spectra $C_{\ell}^{TT}$, $C_{\ell}^{TE}$, $C_{\ell}^{EE}$, $C_{\ell}^{BB}$ for temperature, cross temperature-polarization and $E$ and $B$ modes polarization using the publicly available Boltzmann code \texttt{camb}~\cite{Lewis:1999bs}. The assumed instrumental noise reads as

\begin{equation}
N_\ell = w^{-1}\exp(\ell(\ell+1)\theta^2/8\ln2)~,
\end{equation}
where $w^{-1}$ is experimental sensitivity expressed in $(\mu K$-rad$)^2$ and $\theta$ is the experimental FWHM angular resolution of the beam. The total variance of the multipoles $a_{\ell m}$ will be given by the sum of the fiducial $C_\ell$'s plus the instrumental noise $N_\ell$. The simulated Planck data also has a contribution from experimental noise similar to that presented in the 2018 Planck legacy release analyses~\cite{Akrami:2018vks}. 

Concerning future CMB observations, we shall consider two future CMB experiments, PICO~\cite{Hanany:2019lle} and PRISM~\cite{Delabrouille:2019thj}, and we shall generate the noise spectra with the noise properties shown in Tab.~\ref{tab:specifications}. 
We also simulate BAO data, computing the mean values of the BAO observables using the $\Lambda$CDM fiducial 
cosmology assumed in this analysis, and taking the uncertainties and covariance matrices of the original BAO data listed in Sec. 5.1 of the Planck parameters paper~\cite{Aghanim:2018eyx}, to build the likelihood. 

We perform Monte Carlo Markov chain (MCMC) analyses to current and future mock data (which are generated assuming the minimal $\Lambda$CDM scenario) assuming a non-zero coupling $\xi$ between the dark matter and the dark energy fluids. In particular, we consider a class of models where the Dark Matter and Dark Energy continuity equations are coupled as follows:
\begin{eqnarray}
\dot{\rho}_c+3{\cal H}\rho_c &=& Q\,, \\
\dot{\rho}_x+3{\cal H}(1+w)\rho_x &=&-Q\,,
\end{eqnarray}
where the dot corresponds to the derivative with respect to conformal time $\tau$, $\mathcal{H}$ is the conformal expansion rate of the universe, $\rho_c$ and $\rho_x$ are the dark matter and dark energy mass energy densities respectively, the dark energy equation of state $w$ is assumed to be constant, and the coupling function $Q$ governing the interaction rate between the two dark components is given by:
\begin{eqnarray}
Q = \xi{\cal H}\rho_x\,,
\label{model}
\end{eqnarray}
\noindent In order to derive the cosmological constraints, we shall compare the theoretical spectra with the mock datasets, considering a Gaussian likelihood ${\cal L}$ given by 

\begin{equation}
 - 2 \ln {\cal L} = \sum_{\ell} (2\ell+1) f_{\rm sky} \left(
\frac{D}{|\bar{C}|} + \ln{\frac{|\bar{C}|}{|\hat{C}|}} - 3 \right),
\label{chieff}
\end{equation}
where $\bar{C}$ and $\hat{C}$ are the assumed fiducial and theoretical plus noise power spectra, $D$ is:
\begin{eqnarray}
D  &=&
\hat{C}_\ell^{TT}\bar{C}_\ell^{EE}\bar{C}_\ell^{BB} +
\bar{C}_\ell^{TT}\hat{C}_\ell^{EE}\bar{C}_\ell^{BB} +
\bar{C}_\ell^{TT}\bar{C}_\ell^{EE}\hat{C}_\ell^{BB} \nonumber\\
&&- \bar{C}_\ell^{TE}\left(\bar{C}_\ell^{TE}\hat{C}_\ell^{BB} +
2\hat{C}_\ell^{TE}\bar{C}_\ell^{BB} \right). \nonumber\\
\end{eqnarray}
and $f_{sky}$ is the sky fraction measured by the experiment (see Refs.~\cite{Capparelli:2017tyx,DiValentino:2016foa,Hanany:2019lle,Delabrouille:2019thj} for more details).

For our numerical analysis, we make use of both the original version and a modified version with the IDE scenario of the publicly available MCMC code \texttt{CosmoMC}~\cite{Lewis:2002ah} package (see \url{http://cosmologist.info/cosmomc/}), implementing an efficient sampling of the posterior distribution using the fast/slow parameter decorrelations \cite{Lewis:2013hha}, and with a convergence diagnostic based on the Gelman-Rubin statistics~\cite{Gelman:1992zz}.

\section{Results}
\label{results}

Table~\ref{table_IDE} presents the constraints at 68\% CL on the seven varying cosmological parameters of the IDE model with a dark energy equation of state $w=-0.999$.
Table~\ref{table_LCDM} illustrates the constraints on the fiducial parameters after fitting the mock data generated with a dark energy equation of state $w=-0.999$ to the $\Lambda$CDM case (i.e. $w=-1$, corresponding to the fiducial model) and therefore always neglecting the presence of a possible interaction rate $\xi$ among the dark sectors. Therefore, the results depicted in Tab.~\ref{table_LCDM} allow us to quantify the error introduced when considering $w$ slightly different from $-1$, i.e. the value assumed in the fiducial cosmology, and serve us as a direct test of the reliability of our method. Notice that we can recover with an exquisite precision the fiducial values of the cosmological parameters chosen to create the mock datasets with all the current and future cosmological observations considered in this work. These results ensure the robustness of our approach and warrant the strength of the derived conclusions, excluding the presence of spurious biases.

\begin{figure}
\includegraphics[width=\columnwidth]{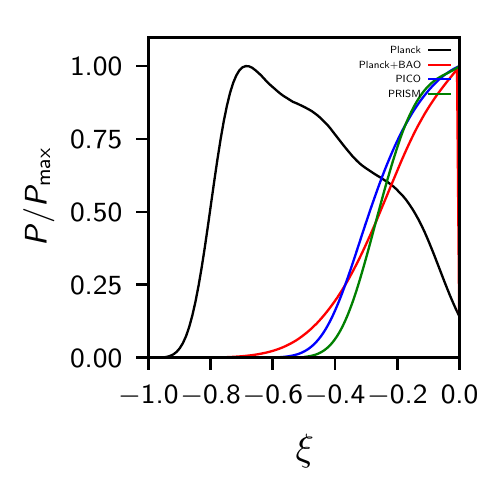}
\caption{One-dimensional marginalized posterior distributions for the parameter $\xi$ governing the dark matter-dark energy interactions, from different data combinations arising from our MCMC analyses, as reported in Table~\ref{table_IDE}, see text for details.}
\label{fig-xi}
\end{figure}

Correlations between the cosmological parameters play a crucial role when exploring exotic scenarios. Indeed, it is well known that the geometric degeneracy present for the IDE models in the CMB damping tail between the matter mass-energy density and the Hubble constant it is a straightforward solution to ease the $H_0$ tension. CMB observations constrain the quantity $\Omega_m h^2$ using the position of the acoustic peaks, and therefore a larger value of $H_0$ can be easily obtained by means of a lower value of $\Omega_m$, which is precisely what happens within the IDE models considered here (see Eq.~(\ref{model})) when the coupling satisfies the condition $\xi<0$, due to the fact that the energy flows from the dark matter sector to the dark energy one. Therefore, parameter degeneracies may potentially lead to fake indications for exotic physics. If we assume that nature has chosen the minimal $\Lambda$CDM scenario, but the \emph{observational} data analysis is performed assuming an IDE Model, i.e. we consider the coupling $\xi$ free to vary when we analyse the mock datasets, we find that a Planck-like experiment (i.e.~cosmic variance limited in temperature but not in polarization) is not powerful enough to recover the fiducial cosmological model in this case. Note from the results shown in Tab.~\ref{table_IDE}, that due to the strong correlation present between the standard and the exotic physics parameters, an evidence at more than $3\sigma$ for a coupling between dark matter and dark energy different from zero is found, i.e. $-0.85<\xi<-0.02$ at 99\% CL. This strong significance for the presence of a coupling leads to a corresponding reduction in the current cold dark matter mass energy density estimate and an increase of $\theta$, the angular size of the horizon at decoupling. Therefore, the \emph{fake} detection, at more than 99\% CL, for a dark matter-dark energy exchange rate, see the black curve in Fig.~\ref{fig-xi}, is completely due to the cosmological parameter degeneracies. This outcome strongly underlines the importance of confronting the results arising from the data analyses with those obtained with simulations. The inclusion of BAO data, a mock dataset built using the same fiducial cosmological model than that of the CMB, helps in breaking the degeneracy, providing a lower limit for the coupling $\xi$ in perfect agreement with zero, see the fourth column of Tab.~\ref{table_IDE} and the red curve in Fig.~\ref{fig-xi}. 
 
Future observations from PICO or PRISM, cosmic variance-limited polarization CMB experiments, will be able to break the correlations between the parameters without the addition of any external datasets. These future observations will be able to  recover (alone) the true, \emph{nature} fiducial cosmology within one standard deviation, see Tab.~\ref{table_IDE} and the green and blue curves in Fig.~\ref{fig-xi}.

\begin{table*}
\centering
\caption{68\% CL bounds on the cosmological parameters assuming that nature has chosen the minimal $\Lambda$CDM scenario but the \emph{observational} data analysis is performed assuming an IDE Model with a dark energy equation of state $w=-0.999$.}
\label{table_IDE}
\resizebox{\textwidth}{!}{  
\begin{tabular}{ c |c c c c c c c} 
  \hline
 \hline                                          
Parameters & Fiducial model & Planck & Planck+BAO & PICO & PRISM \\ \hline

$\Omega_b h^2$ & $    0.02236$ & $    0.02238\pm 0.00015$ & $    0.02230 \pm 0.00014$   &    $    0.022364 \pm 0.000029$  & $    0.022361\pm0.000019$ \\

$\Omega_c h^2$ & $    0.1202$ & $    0.056^{+0.025}_{-0.047}$  & $    0.101^{+0.019}_{-0.006}$ &   $    0.100^{+0.019}_{-0.008}$  & $    0.103^{+0.016}_{-0.007}$  \\

$100\theta_{MC}$ & $ 1.04090$ & $    1.0451^{+0.0021}_{-0.0032}$  & $    1.0419^{+0.0005}_{-0.0011}$  &  $    1.04206^{+0.0005}_{-0.0011}$  & $    1.04191^{+0.00042}_{-0.00094}$ \\

$\tau$ & $ 0.0544$ & $    0.0528^{+0.010}_{-0.009}$  & $    0.0517\pm0.0098$   & $    0.0543^{+0.0016}_{-0.0019}$   & $    0.0542^{+0.0017}_{-0.0019}$ \\

$n_s$ & $ 0.9649$ & $    0.9652\pm 0.0041$ & $    0.9624\pm 0.0036$  &   $    0.9571\pm0.0014$  & $    0.9657\pm0.0012$  \\

${\rm{ln}}(10^{10} A_s)$ & $  3.045$ & $    3.041^{+0.020}_{-0.018}$  & $    3.042\pm 0.019$   &  $    3.0436^{+0.0030}_{-0.0034}$   & $    3.0435\pm0.0032$ \\

$\xi$ & $    0$ &  $    -0.48^{+0.16}_{-0.30}$  & $    >-0.223$  & $    >-0.220$  & $    >-0.195$ \\

\hline \hline 

\end{tabular}
}
\end{table*}

\begin{table*}
\centering
\caption{68\% CL bounds on the cosmological parameters assuming that nature has chosen the minimal $\Lambda$CDM scenario (with $w=-1$) but the \emph{observational} data analysis is perfomed with a dark energy equation of state $w=-0.999$. A zero interaction between the dark sectors has always been assumed.} 
\label{table_LCDM}
\resizebox{\textwidth}{!}{  
\begin{tabular}{ c |c c c c c c c} 
  \hline
 \hline                                          
Parameters & Fiducial model & Planck & Planck+BAO & PICO & PRISM \\ \hline

$\Omega_b h^2$ & $    0.02236$ & $    0.02236\pm 0.00015$ & $    0.02235 \pm 0.00014$   &    $    0.022359 \pm 0.000029$  & $    0.022360\pm0.000020$ \\

$\Omega_c h^2$ & $    0.1202$ & $    0.1201\pm0.0016$  & $    0.1205\pm0.0011$ &   $    0.12020\pm0.00023$  & $    0.12019\pm0.00021$  \\

$100\theta_{MC}$ & $ 1.04090$ & $    1.04091\pm0.00037$  & $    1.04087\pm0.00033$  &  $    1.040897\pm 0.000068$  & $    1.040901\pm0.000060$ \\

$\tau$ & $ 0.0544$ & $    0.054\pm 0.010$  & $    0.0539\pm0.0095$   & $    0.0544\pm0.0018$   & $    0.0546\pm0.0018$ \\

$n_s$ & $ 0.9649$ & $    0.9650\pm 0.0042$ & $    0.9644\pm 0.0033$  &   $    0.9649\pm0.0013$  & $    0.9649\pm0.0011$  \\

${\rm{ln}}(10^{10} A_s)$ & $  3.045$ & $    3.044\pm 0.019$  & $    3.045\pm 0.019$   &  $    3.0451\pm0.0032$   & $    3.0453\pm0.0030$ \\

\hline \hline 

\end{tabular}
}
\end{table*}

\section{Conclusions}
\label{conclu}

Side by side to the early expansion history modifications, that are not efficient in resolving completely the $H_0$ tension, and the late-time solutions, that solving the Hubble constant problem, can not reconcile instead the BAO data, there are the interacting Dark Matter-Dark Energy models. We have studied in this paper whether the constraints obtained for these interacting models derived from current observations, see for example~\cite{DiValentino:2019jae,DiValentino:2019ffd}, are fully supported by simulated Planck data. Intriguingly, we have found here that due to the correlation between the parameters, there exists a dangerous \emph{fake} detection for a non-zero Dark Matter-Dark Energy coupling at many standard deviations when dealing with CMB observations from the Planck satellite.~\footnote{It is worthwhile to note here that, using exactly the same Planck simulated data in~\cite{DiValentino:2019qzk}, it is shown in Fig.1 that a Planck-like experiment is able to constrain $\Omega_k=0$ with a 2\% uncertainty, without a "fake detection" of a closed models, if flat is the fiducial cosmology.}
As a second step, we have tested the same hypothesis exploiting simulations for future cosmic variance-limited polarization CMB experiments, such as PICO or PRISM. We have found that these experiments could be able to break the existing parameter correlations, providing reliable constraints on the cosmological parameters. While these results are obtained for a given interacting model with an energy exchange rate proportional to the dark energy energy density and other models could not lead to the very same fake detection effect,~\footnote{A similar simulated study applied to interacting models with the energy exchange rate $\propto \rho_c$ will be carried out elsewhere.} the main end of our study is to emphasize the utmost importance of confronting the results arising from data analyses to those obtained with simulations before deriving any final conclusions concerning the scrutinized cosmological model, rather than focus on a particular scenario.

\section*{Acknowledgements}
We thank Alessandro Melchiorri and Sunny Vagnozzi for the enlightening discussions. EDV was supported from the European Research Council in the form of a Consolidator Grant 
with number 681431. O.M. is supported by the Spanish grants FPA2017-85985-P, PROMETEO/2019/083 and by the European ITN project HIDDeN (H2020-MSCA-ITN-2019//860881-HIDDeN).

\section*{Data Availability}

The simulated data underlying this article will be shared on reasonable request to the corresponding author.


\bibliographystyle{mnras}
\bibliography{H0}

\bsp	
\label{lastpage}
\end{document}